**Phase-field simulations opening new horizons in corrosion research**

Emilio Martínez-Pañeda

Department of Engineering Science, University of Oxford, Oxford OX1 3PJ, UK

**Abstract**

This work overviews a new, recent success of phase-field modelling: its application to predicting the evolution of the corrosion front and the associated structural integrity challenges. Despite its important implications for society, predicting corrosion damage has been an elusive goal for scientists and engineers. The application of phase-field modelling to corrosion not only enables tracking the electrolyte-metal interface but also provides an avenue to explicitly simulate the underlying mesoscale physical processes. This lays the grounds for developing the first generation of mechanistic corrosion models, which can capture key phenomena such as film rupture and repassivation, the transition from activation- to diffusion-controlled corrosion, interactions with mechanical fields, microstructural and electrochemical effects, intergranular corrosion, material biodegradation, and the interplay with other environmentally-assisted damage phenomena such as hydrogen embrittlement.

**Keywords:** corrosion; diffusion; durability; environment; mesoscale; modeling

## Introduction

Corrosion, the degradation of materials due to their reaction with the environment, is arguably one of the most longstanding challenges in materials engineering [1]. Despite decades of research, the prediction of corrosion failures continues to be an elusive goal [2], and this comes at a great cost – only in the UK, the *yearly* cost of corrosion exceeds 49 billion pounds. Predicting corrosion failures is thus a highly sought-after endeavour but also a remarkably challenging one [3]. Interpolating laboratory data is of little use and delivering reliable, mechanistic predictions would require the simulation of all the relevant coupled electrical, mechanical and chemical phenomena governing the evolution of the corrosion front. Moreover, modelling the morphology of evolving interfaces, such as the electrolyte-metal one, is a notorious mathematical and computational undertaking as it requires defining moving interfacial boundary conditions and manually adjusting the interface topology with arbitrary criteria when merging or division occurs. Tackling this latter obstacle has been made possible by the emergence of phase-field models. Interfaces are no longer sharp but instead smeared over a finite domain and described by means of an auxiliary field variable, the phase-field order parameter $\phi$, which takes distinct values in each of the phases (e.g., 0 and 1) and exhibits a smooth change between these. This *implicit* description of interfaces brings multiple computational advantages, including (i) defining the interface equation in the entire domain, such that no special treatment of the interface is needed, (ii) the ability to naturally capture topological changes such as merging or divisions of interfaces, and (iii) the possibility of easily combining the interface equation with equations describing other physical phenomena, so-called *multi-physics modelling*. In corrosion, being able to track the morphology of the corrosion front, including features such as pits and cracks, is of utmost importance as the problem is strongly coupled – the shape of the corrosion front depends on the local chemical and mechanical fields, but the local chemical and mechanical fields are themselves dependent on the corrosion front morphology (e.g., stresses increase with pit sharpness and the pH changes drastically from the bulk to the inside of occluded areas such as pits and cracks). As a result, phase-field offers an exciting pathway to combine the modelling of the electro-chemo-mechanical phenomena at play with a computationally compelling strategy for describing the evolution of the corrosion front. In this way, the complicated physical processes that govern corrosion, from local electrolyte behaviour to microstructural anisotropies, can be explicitly simulated, reducing the number of assumptions and providing mechanistic, physically-based predictions. This opportunity is also facilitated by the continuous increase in computer power; it is now possible to simulate numerous concurrent physical processes occurring in both the material and the environment (chemical reactions, ion transport, material deformation and fracture, transport of diluted species, fluid flow) – see **Figure 1**. The differential equations of these physical processes are known, and the research challenge lies in establishing universal coupling relationships. Thus, guided by critical experiments, the combination of phase-field approaches and multi-physics modelling can bring the virtues of computer simulations to the discipline of corrosion.

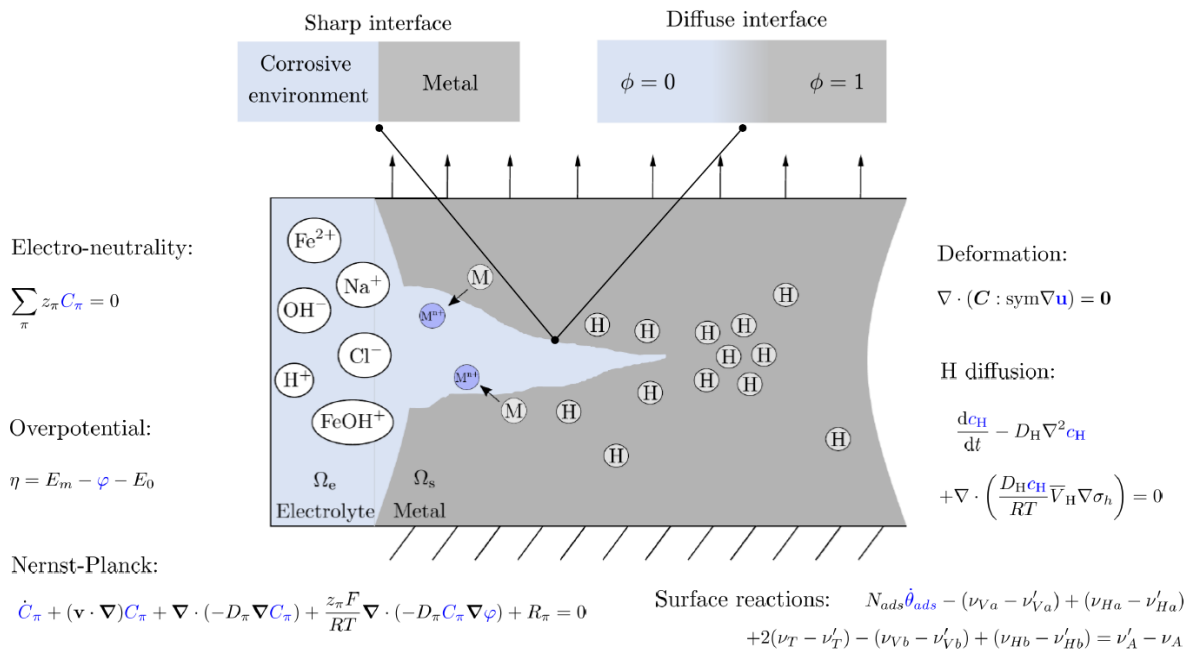

**Figure 1**. The challenge of predicting corrosion failures. Illustration depicting typical phenomena involved in stress corrosion cracking, with the variables that correspond to nodal degrees-of-freedom in a finite element implementation highlighted in blue. The phenomena displayed include the transport of ions due to fluid flow, diffusion and electro-migration, the deformation of the metal, the absorption of hydrogen by the metal, and its transport within the crystal lattice. Material dissolution and the electrolyte-metal interface are described using a diffuse phase-field approach.

The first computational phase-field models for corrosion go back to the mid-to-late 2010s [4–11]. Seminal works include those of Mai and co-workers [6,7,12], showing how phase-field corrosion models can naturally capture both diffusion and activation-controlled corrosion, as well as the transition from one to another. Later, Cui et al. [13] provided a rigorous thermodynamic framework for phase-field corrosion and introduced the interplay with mechanics, extending the model to capture how corrosion kinetics are accelerated by mechanical fields and pioneeringly incorporating the role of film rupture and repassivation. Their model was built upon Gutman's mechanochemical theory [14] and exploited the connection between the current density and the phase-field mobility coefficient to define a mechanics-enhanced mobility coefficient. In addition to the coupling with mechanics, another key extension was to account for the role that the electrochemical behaviour of the electrolyte plays in material dissolution [8,11,12,15–17]. By obtaining numerical solutions for the distribution of relevant ionic species (Nernst-Planck equations) and electrostatic potential (local electroneutrality), the role that variations in concentration and electrostatic potential play on both reaction rates and electrolyte ionic conductivity can be captured. The field has flourished in recent years and phase-field corrosion models have been extended in multiple directions to resolve the complexity of corrosion phenomena [18]. Multi-phase-field formulations have been developed to capture the role that other evolving phases can play in corrosion, such as hydrides [19,20] or insoluble corrosion products [16,21]. The microstructural anisotropy has been accounted for, both at the mechanical and electrochemical levels [22,23]. Makuch et al. have proposed a general boundary condition for the solution potential that enables capturing the formation and charging dynamics of the electric double layer [23]. The interplay with mechanical fracture and other relevant environmental degradation phenomena such as hydrogen embrittlement has also been considered, through novel multi-phase-field approaches [9,24,25]. And phase-field corrosion models are opening new grounds in areas where computational simulations are strongly needed, such as biomaterials where they can enable fast (virtual) experimentation and provide highly-sought 'in-vivo' predictions [26,27]. Examples of these and other applications are provided below, together with a brief description of the theory underlying these models. Emphasis is placed on the generality of the framework, its potential to mechanistically unravel the complexity of corrosion and the challenges and opportunities ahead.

## A phase-field model for corrosion

Let us provide a simple description of the governing equations underlying phase-field modelling efforts. To this end, we consider a system of free energy $F$ and begin by adopting the simplest possible phase-field corrosion model, where only material dissolution and metal ion transport are considered. Later, the role of material deformation and electrochemistry are discussed.

### Fundamentals of phase-field corrosion

Consider a metal exposed to a corroding electrolyte, as per the sketch of **Figure 2**. We define two primary fields: a phase-field order parameter $\phi$, describing the short-range interactions (material dissolution), and a normalised concentration of metallic ions $c$, describing the long-range interactions (metal ions diffusing away from the corrosion front into the electrolyte). The latter is defined as $c = c_M/c_{solid}$, where $c_M$ is the concentration of metal ions and $c_{solid}$ the concentration of ions in the metal, such that $c = 1$ in the metal and then decreases within the electrolyte until reaching $c \approx 0$ in electrolyte locations far away from the corrosion front.

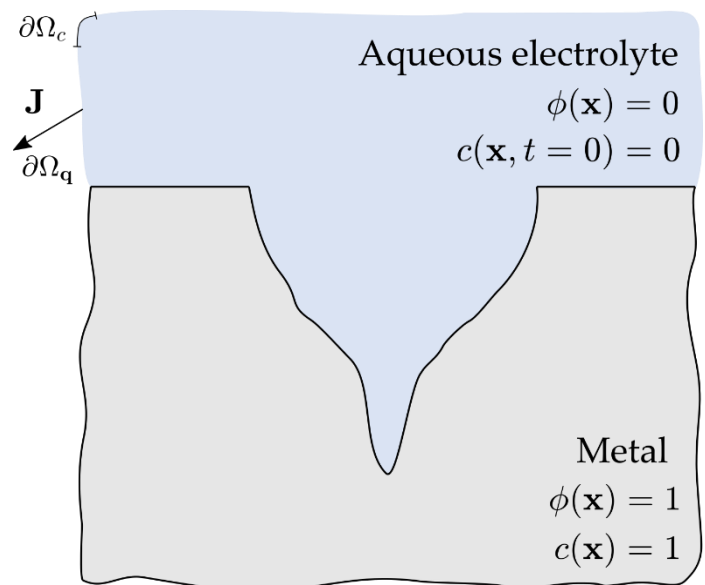

**Figure 2**. A diffusion-phase-field description of the evolution of a corroding metal assuming no mechanical straining and neglecting the influence of electric and electrochemical fields.

As per the usual phase-field formalism [28], $c$ is a conserved field variable constrained by mass balance considerations and $\phi$ is a non-conserved field variable used to indicate which phase is present at a particular location (metal $\phi = 1$ or aqueous electrolyte $\phi = 0$). Noting that the gradient of one of these quantities suffices to characterise the interface contribution, we can write the free energy of the system as:

$$F = F_b + F_i = \int \left( f_c(c, \phi) + f_i(\nabla\phi) \right) \, dV = \int \left( f_c(c, \phi) + \frac{\alpha}{2} (\nabla\phi)^2 \right) \, dV \qquad (1)$$

Where $f_c$ and $f_i$ respectively denote the chemical free energy density and the interface energy density, with the latter being a function of the phase-field gradient and the gradient energy coefficient $\alpha$.

The balance equations are then obtained by minimising the free energy of the system $F$, such that,

$$\frac{\partial\phi}{\partial t}(\mathbf{x}, t) = -L \frac{\delta F}{\delta \phi} = -L \left( \frac{\partial f_c}{\partial \phi} - \alpha \nabla^2 \phi \right) \qquad (2)$$

$$\frac{\partial c}{\partial t}(\mathbf{x}, t) = \nabla \cdot M \nabla \frac{\delta F}{\delta c} = \nabla \cdot M \left( \nabla \frac{\partial f_c}{\partial c} \right) \qquad (3)$$

Where $L$ is the so-called phase-field mobility parameter, and $M$ is the diffusion mobility coefficient, which can be re-written as $M = D/(2A)$, where $D$ is the diffusion coefficient and $A$ is a temperature-dependent free energy density proportionality constant, also known as the free energy density curvature. Equations 2 and 3 are often referred to as the Allen-Cahn (relaxation) and Cahn-Hilliard (diffusion) equations, with the latter here devoid of a gradient term. The chemical free energy density $f_c$ can then be defined considering that the concentration at any point is evaluated as the weighted sum of the solid and liquid concentrations. Accordingly, for a double well potential $g(\phi) = \phi^2(1 - \phi)^2$ with weight $w$,

$$f_c(c, \phi) = h(\phi) f_S(c_S) + \left( 1 - h(\phi) \right) f_L(c_L) + w g(\phi) \qquad (4)$$

where $h(\phi)$ is an interpolation function that must satisfy $h(0) = 0$ and $h(1) = 1$ and $h'(0) = h'(1) = 0$; a common choice is $h(\phi) = -2\phi^3 + 3\phi^2$ [6,15,29]. As discussed in the Appendix of Ref. [15], the height $w$ and gradient energy coefficient $\alpha$ can be related to the interface energy per area $\gamma$ and its thickness $\ell$ as:

$$\gamma = \sqrt{\frac{\alpha w}{18}} \qquad \text{and} \qquad \ell = \sqrt{\frac{8\alpha}{w}} \qquad (5)$$

It remains to define the constitutive choices for the liquid and solid parts of the chemical free energy density, $f_L$ and $f_S$, respectively. A popular approach in the phase-field corrosion community is to adopt the so-called Kim–Kim–Suzuki (KKS) model [30], whereby the material is regarded as a mixture of two coexisting phases at each point and local equilibrium between the two phases is always assumed. Accordingly,

$$c = h(\phi) c_S + \left( 1 - h(\phi) \right) c_L \qquad \text{and} \qquad \frac{\partial f_S(c_S)}{\partial c_S} = \frac{\partial f_L(c_L)}{\partial c_L} \qquad (6)$$

where $c_S$ and $c_L$ respectively denote the normalised concentrations in the solid and the liquid. Assuming a similar $A$ for the solid and liquid phases, the free energy densities read:

$$f_S = A(c_S - c_{Se})^2 = A(c_S - 1)^2 \quad \text{and} \quad f_L = A(c_L - c_{Le})^2 = A\left(c_L - \frac{c_{sat}}{c_{solid}}\right)^2 \qquad (7)$$

where $c_{Se} = c_{solid}/c_{solid} = 1$ and $c_{Le} = c_{sat}/c_{solid}$ are constants denoting the normalised equilibrium equations in the solid and liquid phases. Here, $c_{sat}$ is the saturation concentration, at which the surface concentration of metal ions cannot longer increase due to the precipitation of a chloride salt film and thus corrosion becomes governed by the diffusion of metal ions away from the corrosion front (diffusion-controlled corrosion). Finally, inserting these constitutive choices into equations 2 and 3 renders the following local balance equations:

$$\frac{\partial \phi}{\partial t} = -L\{2A(c_{Le} - 1)[c - h(\phi)(1 - c_{Le}) - c_{Le}]h'(\phi) + wg'(\phi) - \alpha\nabla^2\phi\} \qquad (8)$$

$$\frac{\partial c}{\partial t} = \nabla \cdot D\nabla[c - h(\phi)(1 - c_{Le}) - c_{Le}] \qquad (9)$$

Solving these differential equations numerically enables predicting the evolution of the corrosion front, for arbitrary geometries and dimensions, based on thermodynamic and kinetic principles. Moreover, the model can naturally capture (without 'ad hoc' criteria) the transition from activation-controlled corrosion to diffusion-controlled corrosion; when the normalised metal ion concentration ($c = c_M/c_{solid}$) becomes close to $c_{Le} = c_{sat}/c_{solid}$. The diffusion vs activation controlled corrosion competition is governed by the ratio between the phase-field mobility coefficient $L$ and the diffusion coefficient $D$, which dictates the rate-limiting process.

Therefore, parameters $D$ and $L$ play a key role in the predicted corrosion kinetics. The diffusivity of metal ions is known and on the order of $D = 1 \times 10^{-4}$ mm²/s, taking a value of $D = 8.5 \times 10^{-4}$ mm²/s for steel ions [31]. The phase-field mobility parameter $L$ can be quantitatively related to the corrosion current density $i$, as follows. Considering conditions of activation-controlled corrosion, the velocity of the corrosion front $v_n$ is related to $i$ through Faraday's second law for electrolysis:

$$v_n = \frac{i}{zFc_{solid}} \qquad (10)$$

Here, $z$ is the average charge number ($z = 2.1$ for steels) and $F$ is Faraday's constant. Recall now the phase-field corrosion balance equation, equation 2. Since $v_n \propto d\phi/dt$, the combination of equations 10 and 2 implies that the interface mobility coefficient is proportional to the corrosion current density: $L \propto i$. Accordingly, for any corrosion current density (e.g., $i_a$), the associated mobility coefficient ($L_a$) can be readily determined if the proportionality constant is known for another current (e.g., $L_b/i_b$); i.e.,

$$L_a = i_a \frac{L_b}{i_b} \qquad (11)$$

Therefore, one can run a uniform corrosion simulation with an arbitrary choice of $L$, yet sufficiently small so that it falls within the activation-controlled corrosion regime, compute the corrosion velocity $v_n$ and use equation 10 to estimate the current density associated with that $L$. Once the $L/i$ ratio is known, equation 11 can be used to obtain the interface mobility coefficient for any given corrosion current density. The remaining parameters of the model are the free energy density curvature $A$, the concentration of atoms in the metal $c_{solid}$, the saturation concentration $c_{sat}$, the interface energy per area $\gamma$, and the phase-field length scale $\ell$, with the choice of the last two determining the gradient energy coefficient $\alpha$ and the energy barrier height associated with the double well potential $w$. The concentration quantities are well-established; for the iron-based model system under consideration: $c_{solid} = 143$ mol/L [32] and $c_{sat} = 5.1$ mol/L [33]. The Gibbs (chemical) free energy density is usually approximated with parabolic functions for both liquid and solid phases and a choice of $A = 5.35 \times 10^7$ N/m² yet this choice could arguably be improved by using first-principles calculations or establishing a connection with a thermodynamic database such as CALPHAD [34,35]. The same applies to the interface energy per area, typically taken to be $\gamma = 10$ N/m. It should be noted that the choices of $\gamma$ and one of the phase-field model parameters ($\ell$, $\alpha$, $w$) will determine the other two, as per equation 5. Also, the interface energy $\gamma$ and the gradient energy coefficient $\alpha$ can be related through the barrier of between bulk properties of coexisting phases ($\Delta f_{max}$), a measurable property, as $\gamma = \sqrt{\alpha \Delta f_{max}}$. However, a common practice in the phase-field corrosion community has been so far to set $\ell$ based on numerical considerations, and thus determine $\alpha$ and $w$ from this choice (and that of $\gamma$, which has a physical background).

The aforementioned choices of model parameters have numerical implications, as they make the resulting partial differential equation stiff. Consistent with the above discussion, consider the following standard choices of parameters: $\gamma = 10$ N/mm, $\ell = 5 \times 10^{-6}$ m, $D = 8.5 \times 10^{-10}$ m²/s, $L = 2$ m²/(N s), $A = 5.35 \times 10^7$ N/m², $c_{solid} = 143$ mol/L, and $c_{sat} = 5.1$ mol/L. These lead to $\alpha = 7.5 \times 10^{-5}$ N and $w = 2.4 \times 10^7$ N/m², as per equation 5. Then, the governing equations can be approximated as

$$\frac{\partial \phi}{\partial t} = 2 \times 10^8 [c - 0.04 - 0.96 h(\phi)] h'(\phi) - 4.8 \times 10^7 g'(\phi) + 1.5 \times 10^{-4} \nabla^2 \phi \quad (12)$$

$$\frac{\partial c}{\partial t} = 8.5 \times 10^{-10} \nabla^2 c - 8.5 \times 10^{-10} \nabla^2 [0.96 h(\phi) + 0.04] \quad (13)$$

It can be readily noticed that while the coefficients of the reaction and diffusion terms in equation 13 are in the same magnitude, $O(10^{-10})$, there is significant disparity between the coefficients of the phase-field equation – up to 12 orders of magnitude difference, $O(10^8)$ vs $O(10^{-4})$, making the problem very stiff with respect to $\phi$. This might require the use of small time increments in a backward Euler setting, which has prompted the development of space-time adaptive finite element approaches [36,37]. Numerical convergence can be facilitated by using instead a non-dimensional form of the governing equations [26,38].

*Phase-field corrosion and mechanics*

Let us now consider the interplay with mechanics, as sketched in **Figure 3**. A mechanical free $F_m$, can then be defined, such that the free energy of the system reads,

$$\begin{aligned} F = F_b + F_i + F_m &= \int (f_c(c, \phi) + f_i(\nabla \phi) + f_m(\phi, \boldsymbol{\varepsilon})) \; \mathrm{d}V \\ &= \int \left( f_c(c, \phi) + \frac{\alpha}{2} (\nabla \phi)^2 + h(\phi) \psi(\boldsymbol{\varepsilon}) \right) \mathrm{d}V \end{aligned} \quad (14)$$

Where $f_m$ is the mechanical energy density and $\psi$ is the strain energy density, which is dependent on the strain tensor $\boldsymbol{\varepsilon}$. The interpolation function $h(\phi)$, degrading the strain energy density, has been chosen to be the same as that used in equation 4, for simplicity, but other choices are also possible [39].

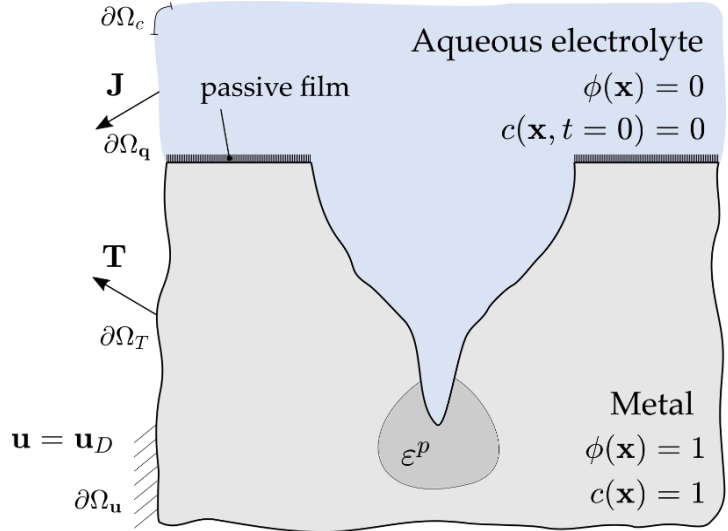

**Figure 3**. A coupled mechanics-diffusion-phase-field description of the evolution of a corroding metal undergoing mechanical straining and partially protected by a passive film.

The strain energy density can be divided into its elastic and plastic counterparts. For example, consider a metal that exhibits power law hardening, as described by the following relationship between the flow stress $\sigma_f$ and the equivalent plastic strain $\varepsilon^p$:

$$\sigma_f = \sigma_y \left(1 + \frac{E\varepsilon^p}{\sigma_y}\right)^N \tag{15}$$

where $\sigma_y$ is the initial yield stress, $E$ is Young's modulus, and $N$ is the strain hardening exponent ($0 < N < 1$). Then, the strain energy density is defined as,

$$\psi = \psi^e + \psi^p = \frac{1}{2}\boldsymbol{\varepsilon}^e : \boldsymbol{C}^e : \boldsymbol{\varepsilon}^e + \frac{\sigma_y^2}{E(N+1)}\left[\left(1 + \frac{E\varepsilon^p}{\sigma_y}\right)^{N+1} - 1\right] \tag{16}$$

With $\boldsymbol{C}^e$ being the fourth-order elastic stiffness tensor. This establishes a two-way coupling between material dissolution and mechanical fields. On one side, the strong form of the mechanical problem becomes,

$$\nabla \cdot [h(\phi)\boldsymbol{\sigma}] = \nabla \cdot [h(\phi)\boldsymbol{C}^{ep}(\boldsymbol{\varepsilon}^e + \boldsymbol{\varepsilon}^p)] = \boldsymbol{0} \tag{17}$$

Where $\boldsymbol{\sigma}$ is the Cauchy stress tensor and $\boldsymbol{C}^{ep}$ is the elastic-plastic material Jacobian. Thus, the coupling defined induces a loss of load-carrying capacity in corroded regions, consistent with observations. On the other side, the mechanics-enhanced phase-field evolution equation reads,

$$\frac{\partial \phi}{\partial t} = -L\{2A(c_{Le} - 1)[c - h(\phi)(1 - c_{Le}) - c_{Le}]h'(\phi) + wg'(\phi) + h'(\phi)\psi(\boldsymbol{\varepsilon}) \\ - \alpha\nabla^2\phi\} \tag{18}$$

Relative to equation 8, there is an additional term $h'(\phi)\psi(\boldsymbol{\varepsilon})$ which accelerates corrosion kinetics with increasing mechanical straining. While there is evidence of corrosion kinetics being enhanced by mechanical fields, this is only observed in activation-controlled corrosion conditions [14] and not relevant to diffusion-controlled corrosion, where material dissolution is only governed by ionic transport. Accordingly, this has led authors to deviate from this variationally-consistent coupling and instead define the interplay with mechanics through the phase-field mobility coefficient $L$. As discussed above, there is a proportional relationship between $L$ and the corrosion current density. Exploiting this, Cui et al. [13] defined the mobility coefficient to be dependent on the equivalent plastic strain and the hydrostatic stress $\sigma_h$ as:

$$L(\sigma_h, \varepsilon^p) = k_m L_0 = \left(\frac{\varepsilon^p}{\varepsilon_y} + 1\right) \exp\left(\frac{\sigma_h V_m}{RT}\right) L_0 \qquad (19)$$

With $L_0$ being the mobility coefficient in the absence of mechanical straining and $k_m$ being a mechanochemical term based on the mechanics-enhanced definition of the corrosion current density put forward by Gutman [14]:

$$i(\sigma_h, \varepsilon^p) = k_m i_0 = \left(\frac{\varepsilon^p}{\varepsilon_y} + 1\right) \exp\left(\frac{\sigma_h V_m}{RT}\right) i_0 \qquad (20)$$

where $V_m$ is the molar volume, $R$ is the gas constant and $T$ is the absolute temperature. This approach has proven to be capable of quantitatively predicting the experimentally-observed enhancement of corrosion kinetics due to mechanical straining [13].

The second key effect that mechanics typically has in corrosion is the disruption of the passive film that often hinders corrosion in a number of metal-electrolyte systems. A classic example are stainless steels, where the presence of chromium leads to the development of a thin film of chromium oxide on the metal surface, which is often termed 'passive' due to its resistance to react with other chemical elements. However, the structural integrity of this nano-meter thick passive film can be readily compromised by mechanical straining, leading to localised corrosion. The effectiveness of the passive film in hindering corrosion is thus dependent on the competition between re-passivation and straining kinetics. This interplay, typically described by the film rupture-dissolution-repassivation (FRDR model) can be incorporated into phase-field corrosion by defining the mobility coefficient as [13]:

$$L = \begin{cases} L_0 & \text{if} \quad 0 < t_i \le t_0 \\ L_0 \exp\left(-k(t_i - t_0)\right) & \text{if} \quad t_0 < t_i \le t_0 + t_f \end{cases} \qquad (21)$$

where $L_0$ is the mobility coefficient associated with the bare metal corrosion current density, $t_0$ is the time that it takes for the repassivated film to start reducing the corrosion current density after a film rupture event, $t_i$ is the total time of each FRDR cycle and $k$ is the parameter that governs the rate of corrosion current decay due to film stabilisation. It remains only to define the film rupture event, which is assumed to take place when the plastic strain accumulated over one FRDR cycle reaches a critical quantity [13].

In addition, phase-field corrosion models have been extended to simulate as well metallic fracture, through the definition of multi-phase-field formulations that can capture the interplay between material dissolution and rupture, as often observed in localised corrosion [24,25]. In these models, the evolution of the crack-intact material interface is described through an additional auxiliary phase-field order parameter, which evolves based on the thermodynamics of fracture [40,41]. The reader is referred to the generalised model by Cui et al. [25] for additional details on the coupling between phase-field corrosion and fracture descriptors. Such a framework enables also capturing the interplay between anodic dissolution-driven and hydrogen-driven stress corrosion cracking mechanisms, as interstitial hydrogen diffusion can be readily modelled and coupled to the damage equation, as done in hydrogen embrittlement phase-field models [42–44].

*Phase-field and electrochemistry*

Finally, we present the structure of phase-field corrosion models that account for electrical and electrochemical effects. For simplicity, as shown in **Figure 4**, the interplay with mechanics discussed in the previous sub-section is not discussed here but one could readily merge both formulations to present a generalised electro-chemo-mechanical theory.

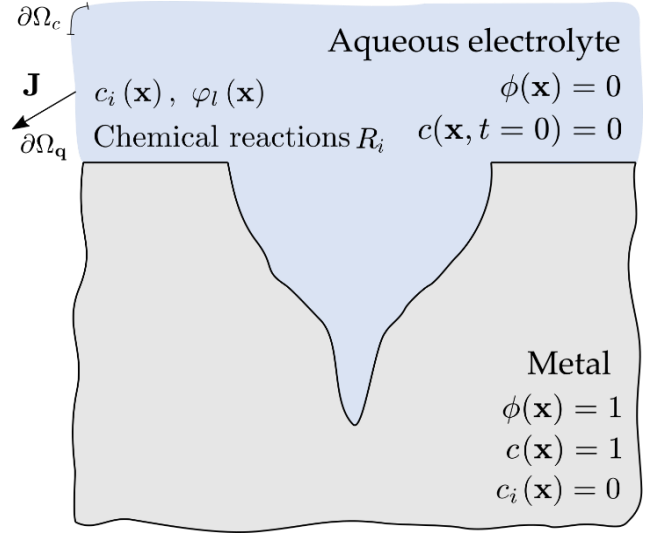

**Figure 4**. A coupled electrochemical diffusion-phase-field description of the evolution of a corroding metal assuming no mechanical straining.

The system free energy now includes a contribution from the electrochemical free energy $F_{el}$ such that,

$$
\begin{aligned}
F = F_b + F_i + F_{el} &= \int \left( f(c,\phi) + f_i(\nabla\phi) + f_{el}(c, c_i, \varphi_l) \right) \, \mathrm{d}V \\
&= \int \left( f(c,\phi) + \frac{\alpha}{2}(\nabla\phi)^2 + F\varphi_l \left( z_M c_{\text{solid}} c + \sum_{i=1}^{n} z_i c_i \right) \right) \mathrm{d}V
\end{aligned}
\tag{22}
$$

where $F$ is Faraday's constant, and $z_M$ and $z_i$ respectively denote the charge number of the metal $M$ and of the $i$th ionic species. Importantly, equation 22 includes two primary fields that come into play when considering the role of electrochemistry: the electrolyte potential $\varphi_l$ and the concentration of as many ionic species as relevant $c_i$, with $i$ being (e.g.) Na+, Cl-, H+, OH-, etc. The former is related to the overpotential $\eta$ as,

$$
\eta = \varphi_s - \varphi_l - E_{eq}
\tag{23}
$$

where $\varphi_s$ is the solid (applied potential) and $E_{eq}$ is the equilibrium potential.

The evolution of the primary fields can then be readily derived (see Ref. [15]). Considering the relationship between the chemical potential, free energy, flux and mass balance, a phase-field-dependent form of the Nernst-Planck equation can be formulated to describe the transport of ionic species as,

$$
\frac{\partial c_i}{\partial t} - \nabla \cdot \{ [1 - h(\phi)] D_i \nabla c_i \} - \nabla \cdot \left\{ \frac{[1 - h(\phi)] D_i c_i}{RT} F z_i \nabla \varphi_l \right\} = R_i
\tag{24}
$$

which incorporates the contributions of diffusion and electro-migration on ionic transport. The phase-field term ensures that transport is limited to the electrolyte (see Ref. [10] for an alternative approach based on smoothed boundary methods). In equation 24, the role of fluid flow has been neglected but this contribution can also be readily incorporated [45]. In equation 24, $D_i$ and $R_i$ respectively denote the diffusion coefficient and the reaction term of the $ith$ ionic species. On the other side, the evolution of the electrolyte potential is given by the following Poisson-type equation:

$$\nabla \cdot (\kappa \nabla \varphi_l) = z_M F c_{solid} \frac{\partial \phi}{\partial t} \tag{25}$$

where the right-hand side of equation 25 accounts for the variation in charge density due to the corrosion reaction, with the term $\partial \phi / \partial t$ capturing the creation of electrons due to the dissolution of the metal electrode [15,17,46]. Most often, the influence of the interfacial double layer is not accounted for due to complexities associated with its modelling, but this has been recently conveniently done by Makuch et al. [23], through the definition of a general boundary condition. In equation 25, $\kappa$ denotes the electric conductivity, which is interpolated between the liquid and solid phases as $\kappa = h(\phi)\kappa_s + [1 - h(\phi)]\kappa_l$, with the liquid conductivity being given by,

$$\kappa_l = \frac{F^2}{RT} \left( c_{solid} c D z_M^2 + \sum_i c_i D_i z_i^2 \right) \tag{26}$$

and $\kappa_s$ typically being assigned a sufficiently large value to ensure a uniform distribution of $\varphi_l$ in the solid phase.

Finally, exploiting the analogy between the corrosion current density $i$ and the phase-field mobility coefficient $L$, the latter can be estimated by a Butler–Volmer-type equation as,

$$L = L_0 \left[ \exp\left( \frac{\alpha_a z_M F \eta}{RT} \right) - \exp\left( \frac{(\alpha_a - 1) z_M F \eta}{RT} \right) \right] \tag{27}$$

with $\alpha_a$ being the anodic charge transfer coefficient. Thus, as evident from equations 23 and 27, the phase-field evolution equation is affected by changes in the electrolyte potential through the relationship between $L, \eta$ and $\varphi_l$. The electrostatic potential also has an effect on the transport of ionic species due to the electromigration term in equation 24. Finally, the calculation of the electrolyte potential is influenced by both the phase-field, as per the right-hand side of equation 25 which accounts for the creation of electrons, and the concentration of ionic species, through their influence on the electrolyte conductivity, equation 26. These couplings highlight the complex phenomena underlying corrosion and the need for coupled electro-chemo-mechanical models to capture them. From a numerical perspective, the dissimilar reaction rates and kinetics of each process bring convergence challenges. However, stabilisation strategies such as lumped integration schemes have been developed that significantly alleviate the resulting numerical oscillations [47].

*Computational implementation*

The balance equations described above can be readily implemented using standard numerical methods. The finite element method has been particularly popular in the phase-field corrosion community, with packages such as COMSOL [15,17], PRISMS [22,48], MATLAB [36], ABAQUS [13,25] and MOOSE [49,50] being employed. In this regard, it is worth noting that codes developed in ABAQUS [13], COMSOL [15] and PRISMS [48] have been made freely available to the community[1]. Despite this availability of computational tools, it is important to note that computational cost is one of the obstacles holding back the potential of phase-field research. The finite element mesh needs to be sufficiently fine to resolve the phase-field interface (a characteristic element length 4-5 times smaller than $\ell$) and, as discussed above, the resulting PDEs can be stiff. Therefore, computational developments that can result in more efficient and robust implementations are of utmost importance for the phase-field community.

---

[1] See www.empaneda.com/codes and https://github.com/prisms-center/phaseField

## Applications

The phase-field corrosion models described in the previous Section have been implemented into computational codes and used to gain new, fundamental insight into corrosion behaviour. The following are illustrative examples that showcase the potential of phase-field corrosion models, highlighting the success of phase-field in delivering new fundamental and technological insight. **Figures 5** and **6** illustrate some of these possibilities.

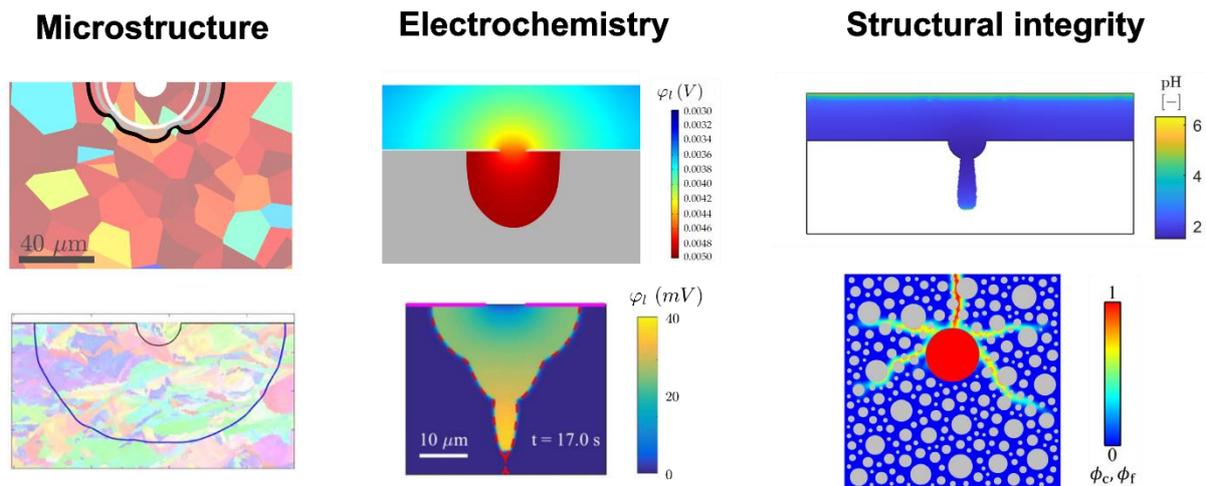

**Figure 5**. Phase-field modelling opening new horizons in corrosion research. Examples of applications: unravelling the role of microstructural anisotropy [22,23], resolving the electrochemistry-corrosion interplay [10,15], and predicting corrosion-induced cracking [24,51].

**Microstructural anisotropy.** There is a need to map the regimes (pit size, environment, material) where the underlying microstructural anisotropy significantly influences corrosion. Phase-field models that can capture the sensitivity of the corrosion potential and mechanical properties to crystallographic orientation have been developed for both additively manufactured [22] and conventional alloys [23]. These not only enable determining the regimes where corrosion is sensitive to the crystallographic anisotropy but can also be directly coupled to EBSD scans to deliver highly accurate predictions and used to design new corrosion-resistant microstructures; the possibilities are vast. Among others, these models have already shown: (i) that variations in the underlying microstructure lead to more extensive defects, faster corrosion kinetics, irregular pit and crack shapes, and highly non-uniform stress and strain states; (ii) that the sensitivity of the corrosion potential to crystallographic orientation plays a much more significant role in dictating the shape and size of corrosion defect than variations in grain size and elastic constants, and (iii) microstructures resulting from additive manufacturing lead to complex and more acute corrosion behaviour.

**Electrochemistry.** Understanding the sensitivity of corrosion kinetics to electrostatic potential and electrolyte characteristics is key to delivering accurate corrosion life predictions. However, this is a remarkable challenge, given the complex interplay between the evolving corrosion defect geometry and local electrochemical conditions (e.g., the pH can drop from a value of 9 in the bulk to a value of 2 inside of a pit [52]). Phase-field corrosion models provide a path for naturally capturing, as an output of the simulation, how the corrosion front morphology and the local electrochemical conditions change in time (see, e.g., [8,15,35,53] and Refs therein). These studies have brought new light into our understanding of corrosion-environment interactions. For example, (i) a strong interplay between corrosion rates and the hydrogen and oxygen evolution reactions has been revealed, (ii) the conditions under which corrosion pits acidify have been quantified, and (iii) how corrosion can become self-sustained in the absence of oxygen [54].

**Structural integrity.** Predicting the pit-to-crack transition has long been a 'holy grail' of the stress corrosion cracking community as it is the phenomenon that starts the damage process that results in catastrophic corrosion failures. The localisation of corrosion defects into sharp pits and cracks can be readily captured by phase-field corrosion models that incorporate the interplay with mechanics [13]. Moreover, multi-field phase-field corrosion models that can capture the interplay between material dissolution and mechanical rupture have been proposed [9,25]. The role that fluid flow, electrochemical transport and reaction kinetics play can be readily modelled [51], as needed to understand stress corrosion cracking and corrosion fatigue phenomena. These models are also key to unravel the longstanding challenge of predicting the durability of reinforced concrete structures [24,54–56]. Importantly, multi-phase-field models for corrosion and fracture have enabled capturing the interplay and competition between anodic dissolution-driven and hydrogen-assisted cracking mechanisms, setting the basis for a generalised model for stress corrosion cracking [25]. Among others, this has revealed that stress corrosion cracking in Al alloys is mostly dominated by the uptake of hydrogen and subsequent embrittlement, with a small range of environmental conditions resulting instead in cracking due to anodic dissolution.

### Insoluble phases     IG corrosion     Biodegradation

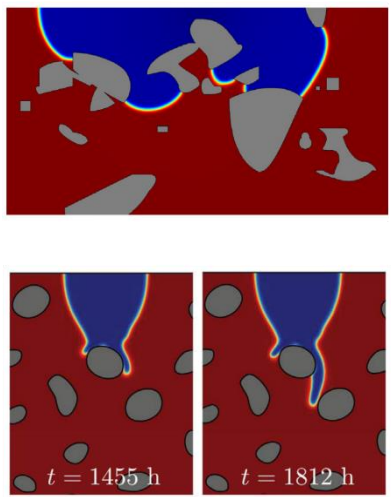
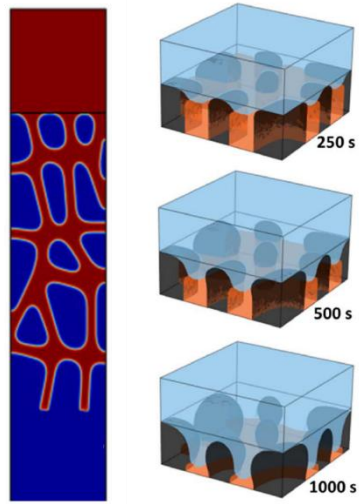
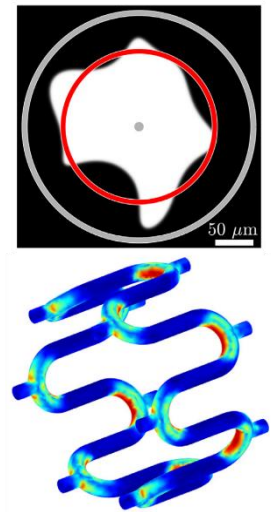

**Figure 6**. Phase-field modelling opening new horizons in corrosion research. Examples of applications: shedding light on the interplay between corrosion and secondary phases [6,7], understanding the preferential path of corrosion along grain boundaries (intergranular corrosion) [57], and predicting 'in vivo' the biodegradation of Mg alloys [26].

**Corrosion-insoluble phases interaction.** A very illustrative example of the complexity of corrosion is its interplay with insoluble phases such as insoluble corrosion products, precipitates, and other insoluble depositions. Phase-field models have shown to be capable of capturing the formation of these insoluble phases, as well as the influence that they can have on corrosion kinetics and the shape of the corrosion front [6,7]. Moreover, the modelling can also show how the morphology of these insoluble depositions evolves in time [21] and associated corrosion phenomena such as galvanic corrosion [12,16]. Among others, phase-field corrosion studies considering not corrodible phases have shown that particulate reinforcements, often embedded in stainless steel to improve wear resistance and strength, can notably change the local stress and strain state, influencing the susceptibility to stress corrosion cracking.

**Intergranular corrosion.** It is often observed that metals exhibit preferential corrosion along sensitized grain boundaries in what is known as intergranular (IG) corrosion. Phase-field models have been developed capable of capturing IG corrosion [21,57] and cracking [9,58], laying the grounds for the design of corrosion-resistant microstructures. These modelling efforts have also resulted in new physical insight. For example, phase-field corrosion models capable of discerning between IG and bulk corrosion have shown how the former dominates in early stages but is slowed down as the transport of metal ions along the narrow, corroded grain boundaries is hindered. These studies have also shed new light on the important role that different heat treatments play and allowed to rationalise how rolled aluminium sheets exhibit IG corrosion that is highly sensitive to the rolling plane direction.

**Material biodegradation (biocorrosion).** There is growing interest in the use of magnesium (Mg) alloys as biomedical implants due to their excellent biocompatibility and ability to naturally biodegrade inside the human body, eliminating the need for a second surgical operation to remove the implant. However, this is being held back by the fast degradation rate of Mg in biological environments, relative to the tissue healing time. Significant experimental research is ongoing aiming at overcoming this by providing a variety of solutions, including alloying, surface treatments and composites design [59,60]. Phase-field models for material biodegradation have recently been developed [26,27]. These can significantly accelerate the assessment of potential solutions and tailor material and component design to target degradation rates. More importantly, 'in vitro' validated electro-chemo-mechanical phase-field biocorrosion models can be used to gain insight of 'in vivo' behaviour [39].

## Concluding remarks

The combination of phase-field and multi-physics modelling has opened new horizons in the prediction of corrosion, a phenomenon known for its complexity and technological importance. While the simulation of corrosion has long been considered a daunting endeavour, phase-field corrosion models have shown to be suitable for the task. The field of phase-field corrosion is still in its very early days but it has already shown its potential to revolutionise the corrosion science community, akin to what phase-field approaches have achieved in other fields. Electro-chemo-mechanical phase-field models have shown to be capable to provide the basis of a virtual platform with the capacity to answer key fundamental questions that have thwarted corrosion scientists for decades, while at the same time being able to deliver predictions for the scales and complex conditions of engineering practice. Exciting times ahead.

## Acknowledgements

The author acknowledges helpful discussions with Dr C. Cui and Dr S. Kovacevic (University of Oxford) and financial support from the UKRI Future Leaders Fellowship [grant MR/V024124/1].

## Declaration of competing interest

The authors declare that they have no known competing financial interests or personal relationships that could have appeared to influence the work reported in this paper.

## References

1.    Raabe, D., Tasan, C. C. & Olivetti, E. A. Strategies for improving the sustainability of structural metals. *Nature* **575**, 64–74 (2019).

2.   Turnbull, A., Horner, D. A. & Connolly, B. J. Challenges in modelling the evolution of stress corrosion cracks from pits. *Eng Fract Mech* **76**, 633–640 (2009).

3.   Martínez-Pañeda, E. Progress and opportunities in modelling environmentally assisted cracking. *RILEM Technical Letters* **6**, 70–77 (2021).

4.   Ståhle, P. & Hansen, E. Phase field modelling of stress corrosion. *Eng Fail Anal* **47**, 241–251 (2015).

5.   Abubakar, A. A., Akhtar, S. S. & Arif, A. F. M. Phase field modeling of V2O5 hot corrosion kinetics in thermal barrier coatings. *Comput Mater Sci* **99**, 105–116 (2015).

6.   Mai, W., Soghrati, S. & Buchheit, R. G. A phase field model for simulating the pitting corrosion. *Corros Sci* **110**, 157–166 (2016).

7.   Mai, W. & Soghrati, S. A phase field model for simulating the stress corrosion cracking initiated from pits. *Corros Sci* **125**, 87–98 (2017).

8.   Ansari, T. Q. *et al.* Phase-field model of pitting corrosion kinetics in metallic materials. *NPJ Comput Mater* **4**, 1–9 (2018).

9.   Nguyen, T. T. *et al.* Modeling of inter- and transgranular stress corrosion crack propagation in polycrystalline material by using phase field method. *J Mech Behav Mater* **26**, 181–191 (2017).

10.  Chadwick, A. F., Stewart, J. A., Enrique, R. A., Du, S. & Thornton, K. Numerical Modeling of Localized Corrosion Using Phase-Field and Smoothed Boundary Methods. *J Electrochem Soc* **165**, C633–C646 (2018).

11.  Tsuyuki, C., Yamanaka, A. & Ogimoto, Y. Phase-field modeling for pH-dependent general and pitting corrosion of iron. *Sci Rep* **8**, 12777 (2018).

12.  Mai, W. & Soghrati, S. New phase field model for simulating galvanic and pitting corrosion processes. *Electrochim Acta* **260**, 290–304 (2018).

13.  Cui, C., Ma, R. & Martínez-Pañeda, E. A phase field formulation for dissolution-driven stress corrosion cracking. *J Mech Phys Solids* **147**, 104254 (2021).

14.  Gutman, E. M. *Mechanochemistry of Materials*. (Cambridge International Science Publishing, Cambridge, UK, 1998).

15.  Cui, C., Ma, R. & Martínez-Pañeda, E. Electro-chemo-mechanical phase field modeling of localized corrosion: theory and COMSOL implementation. *Eng Comput* **39**, 3877–3894 (2023).

16.  Lin, C. & Ruan, H. Multi-phase-field modeling of localized corrosion involving galvanic pitting and mechano-electrochemical coupling. *Corros Sci* **177**, 108900 (2020).

17.  Lin, C., Ruan, H. & Shi, S. Q. Phase field study of mechanico-electrochemical corrosion. *Electrochim Acta* **310**, 240–255 (2019).

18.  Ansari, T. Q., Huang, H. & Shi, S.-Q. Phase field modeling for the morphological and microstructural evolution of metallic materials under environmental attack. *NPJ Comput Mater* **7**, 143 (2021).

19.  Yang, C. *et al.* Hydride corrosion kinetics on metallic surface: A multiphase-field modeling. *Mater Res Express* **8**, 106518 (2021).

20.  Sheng, J. *et al.* A phase-field study on the hydrogen-induced pitting corrosion with solid–solid phase transformation in alpha-Uranium. *Comput Mater Sci* **213**, 111663 (2022).

21.  Ansari, T. Q., Luo, J.-L. & Shi, S.-Q. Modeling the effect of insoluble corrosion products on pitting corrosion kinetics of metals. *Npj Mater Degrad* **3**, 28 (2019).

22.  Brewick, P. T. Simulating Pitting Corrosion in AM 316L Microstructures Through Phase Field Methods and Computational Modeling. *J Electrochem Soc* **169**, 011503 (2022).

23.  Makuch, M., Kovacevic, S., Wenman, M. R. & Martínez-Pañeda, E. A microstructure-sensitive electro-chemo-mechanical phase-field model of pitting and stress corrosion cracking. *(submitted)*.

24.  Fang, X., Pan, Z., Ma, R. & chen, A. A multi-phase-field framework for non-uniform corrosion and corrosion-induced concrete cracking. *Comput Methods Appl Mech Eng* **414**, 116196 (2023).

25. Cui, C., Ma, R. & Martínez-Pañeda, E. A generalised, multi-phase-field theory for dissolution-driven stress corrosion cracking and hydrogen embrittlement. *J Mech Phys Solids* **166**, 104951 (2022).

26. Kovacevic, S., Ali, W., Martínez-Pañeda, E. & LLorca, J. Phase-field modeling of pitting and mechanically-assisted corrosion of Mg alloys for biomedical applications. *Acta Biomater* **164**, 641–658 (2023).

27. Xie, C., Bai, S., Liu, X., Zhang, M. & Du, J. Stress-corrosion coupled damage localization induced by secondary phases in bio-degradable Mg alloys: phase-field modeling. *Journal of Magnesium and Alloys* **12**, 361–383 (2024).

28. Moelans, N., Blanpain, B. & Wollants, P. An introduction to phase-field modeling of microstructure evolution. *CALPHAD* **32**, 268–294 (2008).

29. Abubakar, A. A., Akhtar, S. S. & Arif, A. F. M. Phase field modeling of V2O5 hot corrosion kinetics in thermal barrier coatings. *Comput Mater Sci* **99**, 105–116 (2015).

30. Kim, S. G., Kim, W. T. & Suzuki, T. Phase-field model for binary alloys. *Phys Rev E Stat Phys Plasmas Fluids Relat Interdiscip Topics* **60**, 7186–7197 (1999).

31. Kuo, H. C. & Landolt, D. Rotating disc electrode study of anodic dissolution or iron in concentrated chloride media. *Electrochim Acta* **20**, 393–399 (1975).

32. Scheiner, S. & Hellmich, C. Stable pitting corrosion of stainless steel as diffusion-controlled dissolution process with a sharp moving electrode boundary. *Corros Sci* **49**, 319–346 (2007).

33. Laycock, N. J. & Newman, R. C. Temperature dependence of pitting potentials for stainless steels above their critical pitting temperature. *Corros Sci* **40**, 887–902 (1998).

34. Kumar Thakur, A., Kovacevic, S., Manga, V. R., Deymier, P. A. & Muralidharan, K. A first-principles and CALPHAD-assisted phase-field model for microstructure evolution: Application to Mo-V binary alloy systems. *Mater Des* **235**, (2023).

35. Chen, Q., Li, Z., Tang, S., Liu, W. & Ma, Y. A New Multi-Phase Field Model for the Electrochemical Corrosion of Aluminum Alloys. *Adv Theory Simul* **5**, 2200299 (2022).

36. Gao, H., Ju, L., Li, X. & Duddu, R. A space-time adaptive finite element method with exponential time integrator for the phase field model of pitting corrosion. *J Comput Phys* **406**, 109191 (2020).

37. Gao, H., Ju, L., Duddu, R. & Li, H. An efficient second-order linear scheme for the phase field model of corrosive dissolution. *J Comput Appl Math* **367**, 112472 (2020).

38. Li, B., Xing, H. & Jing, H. New diffusive interface model for pitting corrosion. *Npj Mater Degrad* **7**, 84 (2023).

39. Kovacevic, S., Ali, W., Martínez-Pañeda, E. & LLorca, J. Phase-field modeling of mechano-electrochemical-coupled corrosion of bioabsorbable Mg alloys for biomedical applications. *(submitted)*.

40. Bourdin, B., Francfort, G. A. & Marigo, J. J. *The Variational Approach to Fracture*. (Springer Netherlands, 2008).

41. Kristensen, P. K., Niordson, C. F. & Martínez-Pañeda, E. An assessment of phase field fracture: crack initiation and growth. *Philosophical Transactions of the Royal Society A: Mathematical, Physical and Engineering Sciences* **379**, 20210021 (2021).

42. Martínez-Pañeda, E., Golahmar, A. & Niordson, C. F. A phase field formulation for hydrogen assisted cracking. *Comput Methods Appl Mech Eng* **342**, 742–761 (2018).

43. Kristensen, P. K., Niordson, C. F. & Martínez-Pañeda, E. A phase field model for elastic-gradient-plastic solids undergoing hydrogen embrittlement. *J Mech Phys Solids* **143**, 104093 (2020).

44. Duda, F. P., Ciarbonetti, A., Toro, S. & Huespe, A. E. A phase-field model for solute-assisted brittle fracture in elastic-plastic solids. *Int J Plast* **102**, 16–40 (2018).

45. Hageman, T. & Martínez-Pañeda, E. An electro-chemo-mechanical framework for predicting hydrogen uptake in metals due to aqueous electrolytes. *Corros Sci* **208**, 110681 (2022).

46. Tsuyuki, C., Yamanaka, A. & Ogimoto, Y. Phase-field modeling for pH-dependent general and pitting corrosion of iron. *Sci Rep* **8**, 12777 (2018).

47. Hageman, T. & Martínez-Pañeda, E. Stabilising effects of lumped integration schemes for the simulation of metal-electrolyte reactions. *J Electrochem Soc* **170**, 021511 (2023).

48. Goel, V., Lyu, Y., DeWitt, S., Montiel, D. & Thornton, K. Simulating microgalvanic corrosion in alloys using the PRISMS phase-field framework. *MRS Commun* **12**, 1050–1059 (2022).

49. Sahu, S. & Frankel, G. S. Phase Field Modeling of Crystallographic Corrosion Pits. *J Electrochem Soc* **169**, 020557 (2022).

50. Song, J., Matthew, C., Sangoi, K. & Fu, Y. A phase field model to simulate crack initiation from pitting site in isotropic and anisotropic elastoplastic material. *Model Simul Mat Sci Eng* **31**, 055002 (2023).

51. Hageman, T. & Martínez-Pañeda, E. A phase field-based framework for electro-chemo-mechanical fracture: Crack-contained electrolytes, chemical reactions and stabilisation. *Comput Methods Appl Mech Eng* **415**, 116235 (2023).

52. Mccafferty, E. *Introduction to Corrosion Science*. *Springer* (Springer, 2004).

53. Tantratian, K., Yan, H. & Chen, L. Predicting pitting corrosion behavior in additive manufacturing: electro-chemo-mechanical phase-field model. *Comput Mater Sci* **213**, 111640 (2022).

54. Hageman, T., Andrade, C. & Martínez-Pañeda, E. Corrosion rates under charge-conservation conditions. *Electrochim Acta* **461**, 142624 (2023).

55. Korec, E., Jirasek, M., Wong, H. S. & Martínez-Pañeda, E. A phase-field chemo-mechanical model for corrosion-induced cracking in reinforced concrete. *Constr Build Mater* **393**, 131964 (2023).

56. Korec, E., Jirásek, M., Wong, H. S. & Martínez-Pañeda, E. Phase-field chemo-mechanical modelling of corrosion-induced cracking in reinforced concrete subjected to non-uniform chloride-induced corrosion. *Theoretical and Applied Fracture Mechanics* **129**, 104233 (2024).

57. Ansari, T. Q., Luo, J.-L. & Shi, S.-Q. Multi-Phase-Field Model of Intergranular Corrosion Kinetics in Sensitized Metallic Materials. *J Electrochem Soc* **167**, 061508 (2020).

58. Parks, H. C. W. *et al.* Direct observations of electrochemically induced intergranular cracking in polycrystalline NMC811 particles. *J Mater Chem A Mater* **11**, 21322–21332 (2023).

59. Riaz, U., Shabib, I. & Haider, W. The current trends of Mg alloys in biomedical applications-A review. *J Biomed Mater Res B Appl Biomater* **107**, 1970–1996 (2019).

60. Ali, W., Echeverry-Rendón, M., Kopp, A., González, C. & LLorca, J. Effect of surface modification on interfacial behavior in bioabsorbable magnesium wire reinforced poly-lactic acid polymer composites. *Npj Mater Degrad* **7**, 65 (2023).